\documentclass[conference]{IEEEtran}

\usepackage{amsmath,amsfonts}
\usepackage{algorithm}
\usepackage{adjustbox}
\usepackage{lscape}
\usepackage{siunitx}
\usepackage[noend]{algpseudocode}
\usepackage{textcomp}
\usepackage{fancyvrb}
\usepackage{graphicx}
\usepackage{soul}
\usepackage{booktabs}
\usepackage[utf8]{inputenc}
\usepackage[T1]{fontenc}
\usepackage{rotating}
\usepackage{graphicx}
\usepackage{paralist}
\usepackage{tabularx}
\usepackage{multicol}
\usepackage{multirow}
\usepackage{pbox}
\usepackage{enumitem}	
\usepackage{colortbl}
\usepackage{pifont}
\usepackage{xspace}
\usepackage{url}
\usepackage{tikz}
\usepackage{tabularx}
\usepackage{fontawesome}
\usepackage{lscape}
\usepackage{listings}
\usepackage{color}
\usepackage{showexpl}
\usepackage{anyfontsize}
\usepackage{comment}
\usepackage{soul}
\usepackage{multibib}
\usepackage{multirow}
\usepackage{xcolor}
\usepackage{xspace}
\usepackage{caption}
\usepackage{footnote}
\usepackage{tcolorbox}
\usepackage{booktabs}
\usepackage{fixltx2e}
\usepackage{balance}
\usepackage{amssymb}
\usepackage{ifthen}
\usepackage[normalem]{ulem}

\AtBeginDocument{%
  }

\newboolean{showcomments}

\setboolean{showcomments}{true}

\ifthenelse{\boolean{showcomments}}
{\newcommand{\nb}[2]{
		\fbox{\bfseries\sffamily\scriptsize#1}
		{\sf\small$\blacktriangleright$\textit{#2}$\blacktriangleleft$}
	}
	
}
{\newcommand{\nb}[2]{}
	
}

\newcommand{\ie}{\emph{i.e.,}\xspace}
\newcommand{\eg}{\emph{e.g.,}\xspace}

\newcommand{\etal}{\emph{et~al.}\xspace}
\newcommand{\secref}[1]{Section~\ref{#1}\xspace}
\newcommand{\figref}[1]{Fig.~\ref{#1}\xspace}

\newcommand{\tabref}[1]{Table~\ref{#1}\xspace}

\definecolor{lightergray}{rgb}{0.9,0.9,0.9}
\newtcolorbox{resultbox}{colback=lightergray, arc=0.5mm, top=2mm, bottom=2mm, left=2mm, right=2mm}

\newcommand{\categories}{seven\xspace}
\newcommand{\leaves}{41\xspace}
\newcommand{\sampleSize}{443\xspace}
\newcommand{\numProjects}{54\xspace}
\newcommand{\numInitialProjects}{100\xspace}

\makeatletter 
\newcommand{\linebreakand}{%
\end{@IEEEauthorhalign}
\hfill\mbox{}\par
\mbox{}\hfill\begin{@IEEEauthorhalign}
}
\makeatother 

\begin{document}

\title{A Taxonomy of Self-Admitted Technical Debt \\ in Deep Learning Systems}

\author{\IEEEauthorblockN{Federica Pepe}
	\IEEEauthorblockA{\textit{University of Sannio} \\
		Benevento, Italy \\
		f.pepe8@studenti.unisannio.it}
	\and
	\IEEEauthorblockN{Fiorella Zampetti}
	\IEEEauthorblockA{\textit{University of Sannio} \\
		Benevento, Italy \\
		fiorella.zampetti@unisannio.it}
	\and
	\IEEEauthorblockN{Antonio Mastropaolo}
	\IEEEauthorblockA{\textit{William \& Mary} \\
		Williamsburg, VA, United States\\
		amastropaolo@wm.edu}
	\and
	\linebreakand
	
	\IEEEauthorblockN{Gabriele Bavota}
	\IEEEauthorblockA{\textit{Software Institute}\\
		 \textit{Università della Svizzera Italiana}\\ 
		 Lugano, Switzerland \\
		gabriele.bavota@usi.ch}
	\and
	\IEEEauthorblockN{Massimiliano Di Penta}
	\IEEEauthorblockA{\textit{University of Sannio} \\
	     Benevento, Italy \\
		dipenta@unisannio.it}
}

\maketitle

\maketitle

\begin{abstract}
 The development of Machine Learning (ML)- and, more recently, of Deep Learning (DL)-intensive systems requires suitable choices, \eg in terms of technology, algorithms, and hyper-parameters. Such choices depend on developers' experience, as well as on proper experimentation. Due to limited time availability, developers may adopt suboptimal, sometimes temporary choices, leading to a technical debt (TD) specifically related to the ML code. 
This paper empirically analyzes the presence of Self-Admitted Technical Debt (SATD) in DL systems. After selecting 100 open-source Python projects using popular DL frameworks, we identified SATD from their source comments and created a stratified sample of 443 SATD to analyze manually. We derived a taxonomy of DL-specific SATD through open coding, featuring \categories categories and \leaves leaves.
The identified SATD categories pertain to different aspects of DL models, some of which are technological (\eg due to hardware or libraries) and some related to suboptimal choices in the DL process, model usage, or configuration. Our findings indicate that DL-specific SATD differs from DL bugs found in previous studies, as it typically pertains to suboptimal solutions rather than functional (\eg blocking) problems. Last but not least, we found that state-of-the-art static analysis tools do not help developers avoid such problems, and therefore, specific support is needed to cope with DL-specific SATD.

\end{abstract}




\thispagestyle{empty}


\section{Introduction} \label{sec:intro}

In recent years, Deep Learning (DL) has emerged as a new force that, among other applications, has allowed developers to overcome several limitations of the existing ``shallow'' Machine Learning (ML) techniques, and has tackled problems and challenges that were previously infeasible, creating a real revolution in several industrial sectors. Indeed, DL has been widely used in several domains, such as finance, \eg fraud detection~\cite{alghofaili2020financial}, healthcare to improve diagnosis and treatment of diseases, as well as patient outcomes~\cite{ahmad2023revolutionizing}, autonomous driving~\cite{chib2023recent}, and, last but not least, software engineering~\cite{watson2022systematic}. However, the adoption of DL can be cumbersome and can lead to, especially for business, mission, and safety-critical systems, software bugs producing huge economic losses or even threatening human lives~\cite{tesla,uber}. For this reason, researchers have characterized DL-specific bugs, \ie root causes, symptoms, and affected components, considering the peculiarities of DL systems compared to traditional ones, by looking at two different dimensions, \ie the program level~\cite{humbatova2020taxonomy,meher2024deep}, analyzing production code used to train DL models, and the framework level~\cite{chen2023toward,yang2022comprehensive,zhang2018empirical}, looking at bugs occurring in DL frameworks/libraries used to implement DL systems.

Therefore, ensuring the quality of DL systems is crucial. However, not all problems are necessarily bugs, \ie not all of them cause a ``blocking'' system malfunctioning.  Developers may occasionally leave behind unresolved issues and acknowledge them, \ie creating a Self-Admitted Technical Debt (SATD)~\cite{potdar2014exploratory} highlighting the problem through a source code comment or in the issue tracker~\cite{li2023automatic}. 

There are key differences between bugs and SATD. Bugs represent unintentional errors leading to unexpected or incorrect behavior identified by testing, user reports, and monitoring. When discovered, developers try to fix them, especially for blocking bugs that prevent the system's usage.
SATD, instead, refers to the admittance of suboptimal code or design decisions made by developers for expediency, that are not necessarily related to bugs. While, in conventional code, SATD may be related to readability/maintainability or to minor problems that can occur in unlikely usage scenarios~\cite{zampetti2021self}, in the context of DL systems, the inappropriate selection of a loss function may result in poor model performance.


Characterizing DL-specific SATD is, nevertheless, not as straightforward as it has been done for conventional SATD~\cite{Bavota2016MSR,li2023automatic,potdar2014exploratory}. 
DL-based software, similar to ML-based software, includes extra components not typically found in conventional software, which may introduce new domain-specific challenges and, at the same time, may be the origin of new types of TD unseen in conventional software. For instance, ML/DL software performance is highly dependent on the training data quality, which may represent one reason for SATD occurring in those kinds of systems~\cite{obrien202223}. On the one hand, researchers have characterized SATD in ML applications and tools~\cite{bhatia2023empirical,bogner2021characterizing,obrien202223} highlighting that data pre-processing and model generation components are the most prone to TD. On the other hand, Liu \etal~\cite{liu2020using,liu2021exploratory} looked at several DL frameworks to characterize the presence and removal of SATD. However, to the best of our knowledge, their study is focused on the presence of conventional SATD, and they did not look at new types of DL-specific SATD. 

This study aims to empirically define a \textbf{taxonomy} of DL-specific SATD to design ``ad-hoc'' strategies to trace, manage, and remove them from DL systems. To achieve the aforementioned goal, we performed a hybrid, cooperative card sorting procedure~\cite{spencer2009card} on a statistically significant sample of \sampleSize SATD, which occurred in \numProjects Python open-source projects belonging to an initial sample of \numInitialProjects top-starred Python open-source projects relying on TensorFlow or PyTorch. Such a selection is based on the most popular programming language for ML/DL development and the two most popular DL frameworks, also according to previous software engineering studies~\cite{HanSWDX20}. 
As a result, we obtained a taxonomy of DL-specific SATD, made up of two high-level categories related to infrastructure and DL life-cycle, and featuring a total of \categories categories and \leaves leaves.

The purpose of designing such a taxonomy is manifold. First, knowing the SATD specific to the DL components/code of a system may aid developers to better understand the scenarios within which it is acceptable to live with a poorly performing system. Furthermore, by comparing DL-specific SATD and bugs, it would be possible to provide a set of guidelines aiding developers in identifying whether or not a ``poor'' design decision may result in a crash or failure. Finally, by running static code analysis tools on both DL-specific SATD and SATD that can also occur in conventional software, we checked whether DL developers may rely on existing tools to avoid DL SATD.

\textbf{Replicability:} The dataset and scripts of our study are publicly available for replication purposes~\cite{dataset}. 

\textbf{Structure of the paper.} \secref{sec:design} details the study methodology used to collect SATD from DL systems and to build the taxonomy. The final taxonomy, along with the description of its categories, \eg through examples, is presented in \secref{sec:results}. \secref{sec:discussion} contains a discussion of our findings, with implications for practitioners and researchers, while \secref{sec:threats} reviews the threats to the validity. \secref{sec:related} discusses the related literature, while \secref{sec:conclusion} concludes the paper and outlines directions for future work. 

\section{Study Design} \label{sec:design}

The \emph{goal} of the study is to characterize SATD in open-source DL software. The \emph{quality focus} pertains to understanding SATD types specific to DL (and their differences from DL bugs), with the long-term goal of improving the quality of DL systems.
The \emph{perspective} is of researchers interested in building and improving approaches to identify and fix DL-specific SATD. The study results can also be beneficial to software developers to better understand the nature of SATD involving DL-specific components. The \emph{context} of the study consists of (i) 443 SATDs sampled from 54 (out of an initial set of 100) open-source DL systems (in the following simply referred to as ``projects'') hosted on GitHub, and relying on TensorFlow or PyTorch.
Based on the stated goal, our study aims to address the following research question:

\newcommand{\researchq}{What types of SATD are found in our studied open-source DL systems?}

{\bf RQ: \emph{\researchq}} 

This research question aims to qualitatively define a taxonomy of DL-SATD by discriminating SATD as being specific to DL code/components and those that are generic and
can also occur in conventional systems. Furthermore, we also investigate the similarities and differences between DL SATD and DL bugs. 


\subsection{Data Collection}
The initial step focuses on establishing an appropriate selection of projects for analysis. To this aim, we focus on (i) the Python programming language, which is the most popular language being used for DL applications, and (ii) projects using two DL frameworks, \ie TensorFlow and PyTorch.  These are the two most popular DL frameworks, also investigated in previous software engineering work~\cite{HanSWDX20} along with Theano~\cite{theano}, which, however, we discarded as no longer active. Also, even when one uses other high-level DL frameworks in Python, such as Keras \cite{keras} or HuggingFace transformers~\cite{hugging-face}, TensorFlow or PyTorch are requested.

To identify such projects, we leverage the \textit{Dependents} feature available in the GitHub ``Insights'', \ie given a repository, the dependents page shows a list of all other repositories relying on it. Specifically, we downloaded the dependents of the \textit{tensorflow/tensorflow} and \textit{pytorch/pytorch} GitHub projects.
As there is currently no GitHub API available for analyzing dependents, this task has been performed by leveraging the \textit{Beautiful Soup}~\cite{beautifulsoup:2023} and \textit{requests} Python packages. 
After removing duplicated instances, we collected 112,952 and 160,365 projects, ordered by their number of stars, depending on the TensorFlow and PyTorch libraries, respectively. 

To properly derive a taxonomy of DL-specific SATD, we selected 100 open-source projects from the initial set of candidates. After ordering the projects by stars and excluding forked projects, we manually looked at them to come up with 50 projects relying on TensorFlow and 50 on PyTorch. Our manual analysis aimed to exclude tutorials, code books, toy projects, and projects not primarily written in Python. Furthermore, the number of projects to consider has been determined so that we could have an appropriate number of SATD comments so that we could sample and analyze enough SATD comments for each project.

\subsection{Extracting SATD Comments}
Once identified the projects of interest, we proceeded with the extraction of the SATD comments that can be used to elicit our taxonomy. After cloning each project, we restricted our attention to files importing \texttt{tensorflow} or \texttt{torch}. This was done by analyzing the imports through the Python AST package.

After that, we relied on the \emph{Nirjas}~\cite{nirjas} library to extract all source code comments from such files. Specifically, given a Python file as input, Nirjas provides a JSON file containing information about single-line comments and multi-line comments (comment body together with the line number where the comment begins within the file), other than information about source code elements within it. 
Then, we restricted the set of comments by checking for the presence of an SATD-related keyword. Specifically, we relied on the Keyword-Labeled SATD (KL-SATD) definition provided by Rantala \etal~\cite{rantala2020prevalence}.
We chose to use a lightweight approach for SATD detection instead of relying on ML detection of SATD, \eg using the approach by Ren \etal~\cite{DBLP:journals/tosem/RenXXLWG19}, since our goal was not to collect all possible SATD comments being admitted, but rather, obtain a statistically significant sample to qualitatively analyze, as well as we wanted to limit the number of false positives.
To implement the KL-SATD definition, we relied on regular expressions to identify comments matching one of the following four keywords: \texttt{TODO}, \texttt{FIXME}, \texttt{HACK}, and \texttt{XXX} (all matched as case-insensitive whole words). 
Then, from the comments matching such keywords, we removed those containing nothing but the SATD-related keyword (therefore, not informative).

As a result, we collected a total of 1,306 candidate SATDs across 54 out of 100 initially selected projects. The remaining 46 projects either did not exhibit SATD, or they did not occur in source code files importing TensorFlow or PyTorch. 
Since manually analyzing all candidate SATD comments in our dataset would be time-consuming, we created a randomly stratified sample, where the strata represent the projects, \ie SATD comments have been sampled across projects proportionally to the number of candidate SATD comments from the previous step. 
Given some projects had very few $<5$ SATD candidates, we took at least one SATD comment for each project. This resulted in a sample of 443 candidate SATD comments, guaranteeing a margin of error of $\pm 3.71\%$ with a confidence level of 95\%. The estimation has been performed relying on the sample size ($SS$) formula for an unknown population~\cite{Rosner2011}:
\[
SS=p \cdot \left(1-p\right)\frac{Z_\alpha^2}{E^2}
\]
\noindent and $SS_{adj}$ for a known population $pop$:
\[
SS_{adj}=\frac{SS}{1+\frac{SS-1}{pop}}
\]
\noindent where $p$ is the estimated probability of the observation event to occur, $Z_\alpha$ is the value of the $Z$ distribution for a given confidence level, and $E$ is the estimated margin of error.

\subsection{Taxonomy Construction}

To create a taxonomy of DL-specific SATD, we need to properly identify, among the 443 candidates' SATD, those specific to DL. We consider SATD that may occur in any software system and that were categorized on previous studies on conventional SATD~\cite{Bavota2016MSR,li2023automatic,potdar2014exploratory} as ``not specific''.

To this aim, four authors (henceforth referred to as coders), performed a manual classification of the comments, employing a cooperative hybrid card sorting procedure~\cite{spencer2009card}, \ie multiple people formed the taxonomy collectively and iteratively, with no predefined categories. 
The manual classification was conducted to classify each SATD as (i) \textit{false positive} (\eg \#(todo): $<<name>>$), (ii) \textit{no}: indicating that the SATD is not specific for DL code/components, and (iii) \textit{yes}: the admitted Technical Debt (TD) is specific for DL code/components, and it is unlikely to be annotated in conventional systems. For the latter, the coders provided a label describing the admitted TD type. 
First of all, we conducted a pilot labeling study where coders, in a plenary meeting, classified 30 SATDs in the dataset to ensure a common understanding of the labeling procedure, and to derive the initial set of categories to use for the classification of the remaining ones. After the pilot was completed, the remaining 413 SATD instances were coded incrementally in five rounds involving 50, 100, 100, 100, and 63 SATDs, respectively. 

Within each round, two coders independently validate the SATD assigning one of the previously defined categories or else adding a new one, when needed, using an online spreadsheet displaying the categories defined so far by the coders. The latter has been done to ensure consistent naming without introducing substantial bias. At the end of each round, the remaining two coders checked all the instances considered relevant by at least one of the original coders. It is important to note that we could not estimate the reliability of the study using inter-rater agreement due to the incremental process used to define the DL-specific SATD categories. However, to limit agreement by chance, the two coders not involved in the labeling process discussed and also checked instances where there was an agreement. 

Concerning the saturation, on the one side, we started drafting our taxonomy at the end of the third round. Then, we performed the subsequent rounds to check the extent to which saturation was reached. The  fourth round introduced only two leaves to the taxonomy, \ie ``Inference$\rightarrow$Inappropriate Output Postprocessing'' and ``Training Logic$\rightarrow$Numerical Bug''. The fifth round did not introduce any new category.

\begin{figure*}[h!]
	\centering
	\includegraphics[width=.99\linewidth]{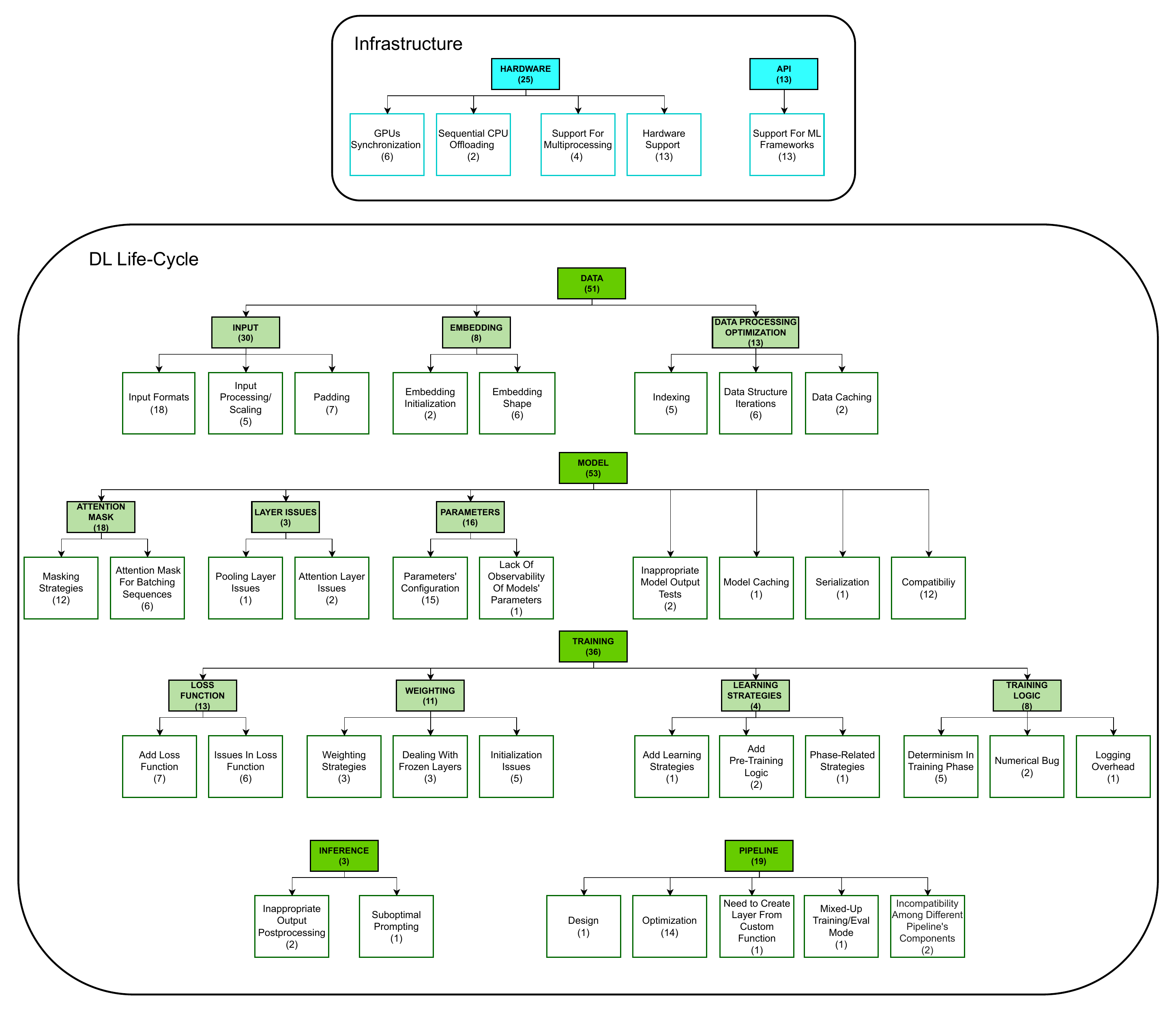}
	\caption{Taxonomy of DL-specific SATD}
	\label{fig:taxonomy}
\end{figure*}

\begin{table*}[ht]
	\caption{List of SATD examples (to open, add \texttt{https://github.com/} before)}
	\label{tab:exampleIdentifiers}
	\centering
	\resizebox{\textwidth}{!}{
	\begin{tabular}{lll} 
		 \hline
		 \textbf{ID} & \textbf{Category} & \textbf{URL} \\   
		 \hline
		 \multicolumn{3}{c}{\sc HARDWARE}  \\ 
		 \hline  
		 E1 & Hardware Support & alpa-projects/alpa/blob/824f2ff/alpa/torch/tensor\_utils.py\#L74 \\   
		 E2 & Support For Multiprocessing & tensorflow/lingvo/blob/cd0a505/lingvo/jax/layers/repeats.py\#L179 \\   
		 E3 & Support For Multiprocessing & huggingface/diffusers/blob/ba59e92f/src/diffusers/pipelines/unidiffuser/pipeline\_unidiffuser.py\#L106 \\ 
		 \hline  
		 \multicolumn{3}{c}{\sc API}  \\  
		 \hline 
		 E4 & Support For ML Frameworks & ivy-llc/ivy/blob/975c8fc/ivy/functional/frontends/tensorflow/general\_functions.py\#L483 \\   
		 E5 & Support For ML Frameworks & microsoft/muzic/blob/5b3890d/emogen/linear\_decoder/command\_seq\_generator.py\#L438 \\   
		 \hline
		 \multicolumn{3}{c}{\sc DATA}  \\   
		 \hline
		 E6 & Input: Input Processing/Scaling & IntelLabs/nlp-architect/blob/88b3236/nlp\_architect/models/tagging.py\#L276 \\   
		 E7 & Input: Input Processing/Scaling & huggingface/diffusers/blob/ba59e92f/examples/community/pipeline\_zero1to3.py\#L667 \\   
		 E8 & Embedding: Embedding Initialization & huggingface/transformers/blob/5936c8c/src/transformers/models/encodec/convert\_encodec\_checkpoint\_to\_pytorch.py\#L230 \\   
		 E9 & Embedding: Embedding Initialization & microsoft/LMOps/blob/909022d/minillm/transformers/src/transformers/models/opt\_parallel/modeling\_opt\_parallel.py\#L856 \\   
		 E10 & Data Processing Optimization: Data Caching & AUTOMATIC1111/stable-diffusion-webui/blob/5ef669d/extensions-builtin/LDSR/sd\_hijack\_ddpm\_v1.py\#L600 \\   
		 E11 & Data Processing Optimization: Indexing & openGVLab/InternGPT/blob/2aceb06/iGPT/models/grit\_src/third\_party/CenterNet2/detectron2/modeling/proposal\_generator/rrpn.py\#L184 \\   
		 E12 & Data Processing Optimization: Data Structure Iterations & huggingface/diffusers/blob/ba59e92f/scripts/convert\_kandinsky\_to\_diffusers.py\#L837 \\   
		 \hline
		 \multicolumn{3}{c}{\sc MODEL}  \\   
		 \hline
		 E3 & Attention Mask: Masking Strategies & flairNLP/flair/blob/603de99/flair/models/lemmatizer\_model.py\#L324 \\   
		 E14 & Attention Mask: Attention Mask For Batching Sequences & alpa-projects/alpa/blob/824f2ff/examples/llm\_serving/model/wrapper.py\#L575 \\   
		 E15 & Layer Issues: Pooling Layer Issues & apache/tvm/blob/c318fa8/tests/python/frontend/pytorch/test\_forward.py\#L2533 \\   
		 E16 & Parameters: Parameters’ Configuration & alpa-projects/alpa/blob/824f2ff/alpa/torch/ops/mapping.py\#L121 \\   
		 E17 & Parameters: Lack of Observability of Models’ Parameters & EleutherAI/gpt-neo/blob/23485e3/utils.py\#L178 \\   
		 E18 & Inappropriate Model Output Tests & Sygil-Dev/sygil-webui/blob/91a4eba/ldm/models/diffusion/ddpm.py\#L1249 \\   E19 & Model Caching & Sygil-Dev/sygil-webui/blob/91a4eba/scripts/scn2img.py\#L147 \\   
		 E20 & Serialization & keras-team/keras/blob/eb12517/keras/legacy/saving/json\_utils.py\#L114 \\   
		 E21 & Compatibility & FedML-AI/FedML/blob/e94d88/python/fedml/core/distributed/communication/s3/remote\_storage.py\#L317 \\   
		 E22 & Compatibility & microsoft/LMOps/blob/909022d/structured\_prompting/fairseq-version/fairseq/examples/roberta/wsc/wsc\_task.py\#L56 \\   
		 \hline
		 \multicolumn{3}{c}{\sc TRAINING}  \\   
		 \hline
		 E23 & Loss Function: Issues in Loss Function & microsoft/LMOps/blob/909022d/understand\_icl/fairseq/examples/MMPT/mmpt/losses/nce.py\#L19 \\   
		 E24 & Loss Function: Add Loss Function & huggingface/transformers/blob/5936c8c/src/transformers/models/longt5/modeling\_longt5.py\#L2076 \\   
		 E25 & Weighting: Weighting Strategies & tensorflow/lingvo/blob/cd0a505/lingvo/jax/train.py\#L469 \\   
		 E26 & Weighting: Dealing With Frozen Layers & huggingface/transformers/blob/5936c8c/src/transformers/models/sam/modeling\_tf\_sam.py\#L577 \\   
		 E27 & Learning Strategies: Phase-Related Strategies & microsoft/LMOps/blob/909022d/understand\_icl/fairseq/examples/MMPT/mmpt/evaluators/predictor.py\#L541 \\   
		 E28 & Training Logic: Determinism in Training Phase & openai/shap-e/blob/0b26c72/shap\_e/models/nn/ops.py\#L407 \\   
		 E29 & Training Logic: Numerical Bug & huggingface/transformers/blob/5936c8c/src/transformers/models/speech\_to\_text/modeling\_tf\_speech\_to\_text.py\#L1292 \\   
		 E30 & Training Logic: Logging Overhead & Stability-AI/stablediffusion/blob/cf1d67a/ldm/models/diffusion/ddpm.py\#L1391 \\   
		 \hline
		 \multicolumn{3}{c}{\sc INFERENCE}  \\   
		 \hline
		 E31 & Inappropriate Output PostProcessing & openGVLab/InternGPT/blob/2aceb06/iGPT/models/grit\_src/third\_party/CenterNet2/tools/deploy/export\_model.py\#L98 \\   
		 E32 & Suboptimal Prompting & oobabooga/text-generation-webui/blob/a0d99dc/extensions/openai/completions.py\#L193 \\   
		 \hline
		 \multicolumn{3}{c}{\sc PIPELINE}  \\   
		 \hline
		 E33 & Design & microsoft/LMOps/blob/909022d/minillm/transformers/examples/research\_projects/wav2vec2/run\_asr.py\#L193 \\   
		 E34 & Need to Create Layer from Custom Function & IntelLabs/nlp-architect/blob/88b3236/nlp\_architect/nn/torch/quantization.py\#L65 \\   
		 E35 & Mixed-Up Training/Eval Mode & allenai/allennlp/blob/80fb606/allennlp/interpret/influence\_interpreters/simple\_influence.py\#L164 \\   
		 E36 & Incompatibility Among Different Pipeline’s Components & huggingface/transformers/blob/5936c8c/src/transformers/configuration\_utils.py\#L341 \\    
		 \hline
	 \end{tabular}
 }
 \end{table*}

At the end of the third round, the four coders worked together to derive a first version of the taxonomy consisting of 37 DL-specific SATD, organized into six categories. The initial taxonomy was revised and refined at the end of the fifth round, ending with a taxonomy consisting of \leaves DL-SATD types organized into \categories categories, further grouped into two high-level categories. 



\section{Study Results} \label{sec:results}

As a result of our card sorting procedure, we defined a taxonomy of DL-specific SATD, depicted in \figref{fig:taxonomy}. On each leaf, the figure also reports the number of occurrences from the sample classified into a given category. The upper-level nodes report the sum of their leaves.

As the figure shows, the taxonomy is organized into two high-level categories, related to Infrastructure and DL Life-Cycle, further specialized into \categories categories and \leaves leaves.

In the following, we discuss the taxonomy along the different categories, and report, when useful, some examples. For such examples, we provide traceability in \tabref{tab:exampleIdentifiers} (the full sample traceability is in our replication package \cite{dataset}). Last but not least, for each category, we summarize the main findings and their implications.

\subsection{Infrastructure}
SATD related to infrastructure pertains to two categories: one related to the usage of hardware (\eg GPUs), and another one to the interaction with external APIs for developing the DL framework/application. 

\textbf{Hardware.} This category accounts for SATD dealing with hardware components, \eg GPUs or TPUs, used for training the DL model or during inference once the trained model is deployed in a production environment. In \textsc{alpa-projects/alpa}, during the process of integrating PyTorch, a developer commented on the need to \textit{``add support for CUDA input tensor''} ($E_1$). The application was not designed to cope with GPUs, while CPUs and meta-input tensors were already handled. Furthermore, in the presence of multiple devices, developers may face challenges regarding the parallelization of certain operations and their synchronization. As an example, in \textsc{tensorflow/lingvo}, a developer asks for the possibility of implementing SPMD (Single Program, Multiple Data) having multiple units cooperating in the execution of a program to obtain results faster, \ie \textit{``Configure for spmd''} ($E_2$). When developing DL applications, the speed for training and inference is not always the only performance requirement to look at. Indeed, we found cases where developers must consider the balance between memory usage and speed. As an example, this involves relying on sequential CPU offloading. Indeed, while the latter saves memory consumption, it also slows down the overall inference process since sub-modules are moved to GPU as needed while returned to the CPU as soon as a new module runs, \ie \textit{``support for moving submodules for components with enable\_model\_cpu\_offload''} in \textsc{huggingface/diffusers} ($E_3$). 

\begin{resultbox}
	\textbf{Finding 1 (Hardware):} One of the main components developers need to care about is \emph{design for change} for what concerns the infrastructure,  to easily support different hardware, as well as parallelization and other optimization elements. Also, it might be valuable to give developers automated hints about possible performance smells that may occur in DL-specific code.
\end{resultbox}

\textbf{API.} This category accounts for SATD due to the usage and integration of DL frameworks. Most of the problems are related to features that DL frameworks do not support. As an example, in \textsc{unifyai/ivy}, there exists a need to use negative indexing. Unfortunately, the used DL framework does not accommodate this requirement, \ie \textit{``\#ToDo: find a way around for negative indexing, which torch does not support''} ($E_4$). 
In some cases, the lack of support for a feature in a DL framework is circumvented through a workaround. This is the case of \textsc{microsoft/muzic}, where a developer acknowledges the presence of a workaround to be removed as soon as the DL framework will explicitly deal with that specific condition, \ie \textit{``\#TODO replace `nonzero(as\_tuple=False)` after TorchScript supports it''} ($E_5$).
\begin{resultbox}
	\textbf{Finding 2 (API):} API-related SATD is usually related to unavailable features, and it may require a temporary workaround. To this extent, features to aid developers to automatically open issues to suggest the desired features will be worthwhile, but also specialized code recommenders suggesting possible workarounds. 
\end{resultbox}

\subsection{DL Life-Cycle}
This high-level category embraces SATD related to various aspects of the DL development life-cycle. Specifically, it features TD related to (i) the type of data and how it is shaped for training purposes, (ii) the training process, such as weighting strategies and loss functions, (iii) the inference process in terms of prompting and post-processing of the produced outcome, (iv) the structure and configuration of the trained DL models, and, (v) the DL pipeline as-a-whole. 

\textbf{Data.} This category groups SATD related to the format/shape of the data used to train a model.
This may concern the \emph{Input} format itself and the extent to which such an input has been properly preprocessed. The latter may negatively impact the model's convergence during training, \eg \textit{``\#TODO: implement method to get only valid logits from the model itself''} in \textsc{IntelLabs/nlp-architect} ($E_6$), or else its performance, \eg \textit{``\# todo in original zero123's inference gradio\_new.py, model.encode\_first\_stage() is not scaled''} in \textsc{hugginface/diffusers} ($E_7$). 

Other common concerns are related to the \emph{Embeddings}, \ie input representation capturing meaningful relationships between different input terms. SATD may concern embedding initialization, \eg in \textsc{huggingface/transformers}, a developer introduced a workaround to handle the type of embedding to use during training as it mismatched with the model, \ie \textit{``\#HACK otherwise .embed gets initialized with .embed\_avg too''} ($E_8$). Other problems are related to their shape, as in \textsc{microsoft/LMOps}, where a developer postponed a proper embedding shape consistency check, \ie \textit{``\# TODO: check tied''} ($E_9$). 

Another aspect that is worth mentioning concerns how inappropriate input data management can lead to efficiency problems in the model training phase, hence requiring a \emph{Data Processing Optimization}. As an example, developers felt it appropriate to reduce---through caching---the overhead associated with data loading during the training process, specifically in the presence of large datasets or complex data pre-processing, \eg \textit{``todo load once not every time''} in \textsc{AUTOMATIC1111/stable-diffusion-webui} ($E_{10}$).  
Problems related to indexing ignored data may lead to performance issues too, \eg in \textsc{OpenGVLab/InternGPT}, a developer acknowledged the poor performance resulting from unneeded operations due to the inappropriate handling of the input data, \ie \textit{``\# TODO wasted indexing computation for ignored boxes''} ($E_{11}$).

Last but not least, we also found concerns related to the need for better data padding (\eg add support for padding in multiple directions), or to improve data structure iterations, mainly to improve the overall performance, \eg \textit{``\#TODO maybe document and/or can do more efficiently (build indices in for loop and extract once for each split?)''} in \textsc{hugginface/diffusers} ($E_{12}$).


\begin{resultbox}
	\textbf{Finding 3 (Data):} Developers may need suitable support to check the conformance of input data with the model and, if needed, pre-process or adapt them. Moreover, the way data is read/processed can introduce performance antipatterns, leaving room for optimization.
\end{resultbox}

\textbf{Model.} This category includes SATD concerning the design and setting of the DL model. It is worth mentioning problems faced when defining the DL model architecture in terms of attention masks, layers,  and other relevant parameters. 

\emph{Attention Masks} are binary masks determining which tokens should be attended to or ignored. SATD we found relate to the choice of masking strategies, or to determining attention masks for batch sequences.
In the \textsc{flairNLP/flair} project, we found a comment highlighting the need to check the attention masking process, ensuring that certain elements are ignored, \ie \textit{``mask out vectors that correspond to a dummy symbol (TODO: check attention masking)''} ($E_{13}$). Other attention mask-related SATD concerns the masking for batch sequences, \eg in \textsc{alpa-projects/alpa} a SATD mentions \textit{``\#Skip the check in DistributedPhysicalDeviceMesh::shard\_args for     attention cache. We need this hack because attention\_cache is     a batch var but alpa doesn't implement a fast path for batch vars.''} ($E_{14}$).

SATD related to \emph{Layers} concerns the presence, in the model, of specific types of layers. We found SATD related to the need to add attention layers or pooling layers. Such pooling layers are typically used (\eg in Convolutional Neural Networks) to reduce input sizes among different layers. In the \textsc{apache/tvm} project, we found a SATD related to the need for fixing a potential issue that could occur in certain circumstances because of the lack of proper pooling: \textit{``\#TODO: Fix VGG and AlexNet issues (probably due to pooling)''} ($E_{15}$).

Once a DL model architecture has been defined in terms of layers and their interconnection, a major issue is the inappropriate configuration of the model \emph{Parameters}. In some cases, such as in \textsc{alpa-projects/alpa}, this is likely done to improve both flexibility and re-usability of the code, \ie \textit{``\dots parameterize this! don't assume NCHW format.''} ($E_{16}$). In other cases, the code does not allow to inspect all model parameters, leading to observability and maintainability issues. As an example, in \textsc{EleutherAI/gpt-neo}, a SATD mentions \textit{ ``TODO: how to get un-trainable dim-names too, batch etc.''} ($E_{17}$).

Other model-related SATD concerns (i) inappropriate model output tests, (ii) model caching, (iii) enhancing serialization features, and (iv) compatibility between different DL components/models. Inappropriate model output test concerns the need to handle cases where the model can produce multiple outputs (\eg multiple solutions ranked using beam search). As an example, in \textsc{Sygil-Dev/sygil-webui}, a developer wants to ensure multiple outputs are never produced, otherwise the output shape itself changes, and this may create issues, \ie \textit{``\#todo cant deal with multiple model outputs check this never happens''} ($E_{18}$).  

Model caching simply concerns ensuring a pre-trained model is loaded from a local cache and not downloaded (to save time), \eg in \textsc{Sygil-Dev/sygil-webui} a SATD mentions \textit{``\#todo: load model from pre-trained keras into user .cache folder like transformers lib is doing it.''} ($E_{19}$).

Serialization concerns the program's features to store trained models on files. We found SATD related to enhance this, and, specifically, in \textsc{keras-team/keras}, to give the possibility to define saved parameters for different scopes, a developer states: \textit{``\#TODO(\dots): Add TF SavedModel scope.''} ($E_{20}$). Such an option, as also discussed in a Stack Overflow post\footnote{\url{https://stackoverflow.com/questions/55245813/saving-restoring-weights-under-different-variable-scopes-in-tensorflow}}, would be very useful in the context of transfer learning, to better save models for multiple fine-tuning conducted for different purposes.

Finally, developers highlight problems in guaranteeing the compatibility among different DL components/models, such as the integration of different DL frameworks, \eg \textit{``\#TODO: added torch model to align the Tensorflow parameters from browser''} in \textsc{FedML-AI/FedML} ($E_{21}$), or the need to rely on workarounds to properly integrate pre-trained models due to different choice for handling specific input data and/or parameters, \eg \textit{``\#hack to handle GPT-2 BPE, which includes leading spaces''} in \textsc{microsoft/LMOps} ($E_{22}$).

\begin{resultbox}
	\textbf{Finding 4 (Model):} The problem of model tuning and, in general, of model configuration is a long-standing one for which reusing previously-adopted solutions may not necessarily be the best option. As far as pre-trained models are concerned, challenges developers face in their integration may be mitigated by providing examples of use, as many models hosted on HuggingFace do~\cite{BhatCHLNZKG23}. 
\end{resultbox}

\textbf{Training.} This category considers SATD dealing with the training process in terms of selecting an appropriate loss function, initializing the model parameters, such as weighting strategies, together with adequate learning strategies, or a sub-optimal implementation of the whole training logic.

As regards the usage of the appropriate \emph{Loss Function}, we encountered two different classes of problems. On the one hand, someone may use loss functions without properly adjusting their parameters, \ie keeping their default values, such as \textit{``\#TODO: define temperature.''} in \textsc{microsoft/LMOps} ($E_{23}$). The temperature is a parameter used in DL models to determine the smoothness of the (softmax) output distribution. That is, a low temperature means more deterministic outputs, whereas a high temperature means more variation. This is also a very important parameter at inference time to make the model more coherent with its training data or more ``creative'' (as it has been studied in recent work~\cite{10.1145/3597503.3623306}).
On the other hand, there are cases requiring the implementation of a customized loss function since it has been proven to be more suitable for the scenario at hand. As an example, in \textsc{huggingface/transformers}, a developer acknowledges the need to \textit{``Add z\_loss''} ($E_{24}$), linking this requirement to an external project which, in a similar scenario, is following the same strategy.

Adopting an appropriate \emph{Weighting} strategy is beneficial for several reasons, such as the appropriate handling of the class imbalance problem or the improvement in terms of model generalizability, \ie guaranteeing good performance on unseen data. As an example, in \textsc{tensoflow/lingvo}, a developer admitted the need to obtain the initial random seeds used for initializing the model's parameters directly from the model definition, rather than manually specifying them, \ie \textit{``\#TODO: Retrieve seeds from model definition instead''} ($E_{25}$), likely to ensure consistency and reproducibility. 
Other weighting problems are related to frozen layers.  In some cases, developers are wondering why the weights of a frozen layer receive a (useless) random initialization, like in \textsc{huggingface/transformers}: \textit{``TODO \dots Why is a non-trainable weight randomly initialized?''} ($E_{26}$).
In other cases, developers suggest explicitly freezing a layer to prevent its retraining or fine-tuning, hence avoiding a possible ``forgetting'' problem, \ie loosing knowledge previously acquired.
Finally, weighting problems concern initialization issues, \ie problems occurring when initializing weights from certain (\eg incompatible) sources.

For what concern the \emph{Learning Strategies}, SATD is related to (i) improving learning-related features, such as adding new learners, (ii) adding specific (customized) logic for using pre-trained models (and eventually fine-tuning them), or (iii) adding specific logic for certain phases of the training (\ie either for the forward or the backward phase), \eg in \textsc{microsoft/LMOps} a SATD mentions \textit{``\#TODO special forwarding logic here.''} ($E_{27}$). 

For what concerns the \emph{Training Logic}, it is worth mentioning the need to guarantee determinism within the training process that would ensure both consistency and reproducibility for the trained model, \eg \textit{``\#TODO shuffle deterministically when not self.training''} in \textsc{openai/shap-e} ($E_{28}$). Indeed, DL frameworks provide explicit APIs to initialize the random seed. For example, TensorFlow uses the \texttt{enable\_op\_determinism} API or the \texttt{TF\_DETERMINISTIC\_OPS} environment variable, while Keras uses the \texttt{set\_random\_seed} API to initialize the random seed. 
While a varying random seed may allow for re-shuffling the starting point over multiple experiments, a constant one ensures the reproducibility of the training process.
Other problems related to the training logic concern numerical bugs, \eg due to repeated approximations, like in \textsc{huggingface/transformers}: \textit{``\#TODO (\dots): investigate why Speech2Text has numerical issues in XLA generate.''} ($E_{29}$). 

Finally, we found SATD related to possible overhead introduced by (excessive) logging, like in \textsc{Stability-AI/stablediffusion} where a developer suggests to disable some expensive logs: \textit{``\#TODO: maybe disable if too expensive''} ($E_{30}$). Clearly, TD like this one introduces trade-off decisions, \ie pondering between performance (with limited logging) and training process troubleshooting.

\begin{resultbox}
	\textbf{Finding 5 (Training):} Like model calibration (which requires complex choices), the lack of setting a proper training phase may lead to performance problems or suboptimal results. Sometimes developers, possibly because of the lack of effort to properly customize the training or to experiment with different options, decide to live with that.
\end{resultbox}

\textbf{Inference.} This category includes SATD related to inappropriate post-processing of the outcome generated when running the trained model, as well as using suboptimal prompting. For the former, in the \textsc{openGVLab/InternGPT} project, there is a comment highlighting the lack of a function to integrate the output of the inference step with subsequent processing or actions, \ie \textit{``\dots inference is now missing post[-]processing glue code''} ($E_{31}$). 
For what concerns prompt-related SATD, one example occurred in the \textsc{oobabooga/text-generation-webui} project, reporting the need to improve the prompting for inference purposes, \ie \textit{``\#XXX User: prompt here also''} ($E_{32}$). Determining the proper prompt to be used during inference---\ie the input query to be seeded into the model, possibly indicating information such as the user's role, the task context, and required output format---has already been studied in previous work~\cite{white2023prompt}.
Last but not least, it must be noted that one problem that may affect the inference phase is the setting of the temperature (as we discussed for the training phase) or other hyper-parameters that affect the inference phase (\eg the beam search, or, for some text transformer models, the parameter to penalize repetitions). However, we did not find such instances in our sample.

\begin{resultbox}
	\textbf{Finding 6 (Inference):} Unsurprisingly, developers highlight the need for improvement, especially for what concerns prompt engineering. While there is intense work in this area, development tools could be enhanced to provide recommendations for suitable prompts for different types of inference.
\end{resultbox}

\textbf{Pipeline.} This category groups SATD dealing with the setting of the DL pipeline in terms of its design and optimization. 

First of all, we found design issues affecting the usage of DL-pipelines, \eg the need to handle certain features through the pipeline, and consequently, for a function, to return a pipeline object, as in \textsc{microsoft/LMOps} ($E_{33}$): \textit{``\#TODO(\dots) return a pipeline (e.g., from jiwer) instead? Or rely on branch predictor as is.''} 

Several projects in our dataset faced problems in terms of deciding what has to be moved outside the pipeline, \eg \textit{``Prepare extra step kwargs. TODO: Logic should ideally just be moved out of the pipeline''} likely to improve modularity and maintainability of the source code. In the examples we found, this was mainly related to extra processing steps, that need to be properly factored out as functions, and not in-lined directly as code in the pipeline. In other circumstances, a suggested (and deferred) change concerns creating a predefined layer to be used in the pipeline, as in \textsc{IntelLabs/nlp-architect}: \textit{``\#TODO(\dots) future work, implement a layer that uses this function that gives a more comfortable.''} ($E_{34}$). 

A recurring concern in pipelines is to force the models' mode, \eg \textit{``\#TODO (\dots): should we make sure `model` is in "train" or "eval" mode here?''} in \textsc{allenai/allennlp} ($E_{35}$). In eval mode, a model works differently, \eg it does not use dropout nodes. While the current (default) mode in the code could be the desired one, an explicit set may avoid possible future bugs. 

Finally, some pipeline-related SATD concerns the compatibility across several components. As an example, in the \textsc{huggingface/transformers} project, we found a comment admitting the need to align the configurations between the tokenizer and the models used in the code, \ie \textit{``\dots TODO: eventually tokenizer and models should share the same config''} ($E_{36}$), likely to improve both consistency and compatibility between the tokenizer and the models.


\begin{resultbox}
	\textbf{Finding 7 (Pipeline):} On the one hand, developers may need better support (\eg low-code tools or IDE-integrated recommenders) to ease the composition of different DL pipeline blocks. On the other hand, suboptimal decisions on what should or should not be in the pipeline may lead to the definition of DL pipeline smells and related refactoring actions.
\end{resultbox}


\section{Discussion} \label{sec:discussion}

In the following, we first discuss, based on the results presented in \secref{sec:results}, the emerging differences between DL bugs---studied in previous research \cite{chen2023toward,ho2023empirical,humbatova2020taxonomy,tambon2024silent,zhang2018empirical}---and DL-SATD, and the adequacy of existing static analysis tools to identify DL-SATD. Then, we outline implications for different stakeholders.

\subsection{Differences between DL SATD and DL bugs}
\label{sec:dlbugs}
Our qualitative findings stress the importance of ensuring the quality of DL systems, not only in terms of avoidance of system crashes, mainly due to the presence of ``blocking'' DL bugs, but also in terms of awareness of SATD, describing the presence of unresolved issues that have to be fixed later on in the system. 

At first glance, our SATD taxonomy would appear (by many category names) similar to some DL bug taxonomies, among others the one by Humbatova \etal~\cite{humbatova2020taxonomy}, \ie the leaves have similar naming but different meanings in terms of the type of effect on the system, \eg on the one side, a system fails to accomplish its task, or it crashes, while on the other side, there may be poor performance in terms of training time or convergence. A detailed mapping between our taxonomy and the one by Humbatova \etal is in our replication package~\cite{dataset}.

One key difference between our study and what Humbatova \etal~\cite{humbatova2020taxonomy} did is that we analyzed source code comments (SATD) left by developers during their coding tasks, while they focused on issues, bug fix commits, Stack Overflow discussions, and interviews with practitioners. Issues preventing the system from properly working are unlikely to be admitted in source code comments. Instead, typically one (i) submits an issue, or (ii) asks for help on Q\&A forums by reporting the failing code snippets and describing what is the desired behavior. As an example, problems dealing with the wrong reference to the GPU  device\footnote{\url{https://stackoverflow.com/questions/40726039/tensorflow-cuda-visible-devices-doesnt-seem-to-work}} make the DL-system fail. However, developers might decide to leave with performance issues due to a lack of parallelization of certain operations and their synchronization. 

Furthermore, in DL systems, one possible root cause for both bugs and SATD deals with tensor types and their shapes. While mismatches between tensors' shape/type represent a blocking issue, DL-SATD mainly describes problems negatively impacting model convergence, as well as the overhead of data loading during training. 

Having said that, there are still commonalities among DL SATD and DL bugs in terms of (i) negative impact on software quality, (ii) contributing to software maintenance challenges, and (iii) need to be properly documented even if within different sources, \ie annotations in code and issues. 

\subsection{Adequacy of state-of-the-practice static analysis tools}
Before thinking about DL-specific SATD recommenders and repayment tools, it would be worthwhile to check whether existing state-of-the-art static analysis tools can at least help spot possible DL-specific TD.

Therefore, we investigated whether code snippets affected by DL-specific SATD in our sample contain at least one warning raised by \textit{Prospector}~\cite{prospector}. Prospector is an aggregator of several static analysis tools for Python, including PyLint~\cite{pylint}, Flake8~\cite{flake8}, or the MyPy type checker~\cite{mypy}.
We conjectured that conventional SATD, such as \textit{``// this method is a nightmare''} or \textit{``TODO: refactor this method is too long''}, can be easily identified by Prospector, while comments such as \textit{``\#TODO: Enable multi-device support''} or \textit{``\#TODO(\dots) - attention mask is not used''} are unlikely to be detected by static code analysis tools. To verify our conjecture, we ran Prospector on the Python files containing SATDs in our sample and manually checked its reports to verify whether or not among the issues being raised there exist some that can be used as a proxy to detect the SATD. 

Our manual analysis (detailed results can be found in our replication package~\cite{dataset}) reveals that out of 432 SATDs in our sample, only 21 (of which only 4 are DL-specific) are somehow related to a warning issued by Prospector. For instance, the conventional SATD \textit{``\#todo no isp\_model?} is identifiable by the violation \textbf{Unused argument 'isp\_model'}, or else \textit{``\#TODO (\dots): The current implementation is ugly. Refactor\dots''} can be associated with the \textbf{Too many statements} violation. Looking into DL-specific SATD, we found only two SATD instances somehow related to a Prospector warning. The first one, (repeated in three code snippets), \textit{``\#TODO: don't match quantizer.weight\_proj''} matched a type/signature violation (\textsc{Signature incompatible with supertype}), while \textit{``\#TODO: maybe have a cleaner way to cast the input (from `ImageProcessor` side?)''} relates to a type violation  (\textsc{Cannot assign to a type.})

In conclusion, the results of this investigation indicate that currently-used static analysis tools cannot capture the essence of DL-specific TD, and, rarely help to identify TD in general (for the latter, confirming previous findings by Rantala \etal~\cite{rantala2023keyword}).
This enforces the need for specialized approaches to properly detect and track them during software evolution. 

\subsection{Implications}
\textbf{Educators} should illustrate how to favor design for change in DL-intensive systems. As we found, this may concern, on the one hand, easing the adaptation to different hardware infrastructures, and, on the other hand, making DL pipelines flexible for what concerns the addition/change of their stages.

Furthermore, although students and professionals today can quickly become proficient with new technologies---thanks to the wide amount of formal and informal documentation available and to advanced tools, including those based on Large Language Models, that support software development---there is still a lack of effective guidance for recognizing, understanding, and correcting poor design and coding practices. 

As far as \textbf{Practitioners} are concerned, a key lesson is related to deciding when a problem in a DL-intensive system is a bug and therefore needs to be fixed as soon as possible, or when it is a SATD and, the system can be released by leaving with that. To this extent, developers should employ a suitable evaluation to identify significant (blocking) deviations from the expected behavior from sub-optimal model accuracy or performance problems whose fix can be postponed. Moreover, as it also happens for conventional systems, developers should be prone to admit TD, as previous work suggested that this may not always be the case in the industry~\cite{zampetti2021self}.

A further implication for practitioners, which could also inadvertently affect other stakeholders, is that, similar to certain types of conventional SATD, DL-specific TD may imply trade-off decisions. As an example, just as performance and maintainability can often be conflicting goals in conventional systems, a faster training process in DL systems may come into conflict with other objectives, such as the ability to effectively monitor the training process (\eg through logs) and, ultimately, the effectiveness of the trained models.

Similarly to what has been done for conventional systems~\cite{DBLP:journals/tosem/RenXXLWG19}, \textbf{Researchers} should develop approaches to (i) identify the presence of ``DL-specific'' bad smells in the source code, and recommend refactoring actions, or else the admittance of a TD, or (ii) propose solutions to repay DL-specific TD. 


\section{Threats to Validity} \label{sec:threats}

\textbf{Construct validity} threats concern the relationship between theory and observation. A possible threat is related to the subjectiveness of our analysis, as the produced taxonomy is based on our interpretation of the SATD comments. We mitigated this threat by having two coders who inspected each SATD candidate, and two more coders who checked all the previously-coded categorizations and resolved the controversial cases.

\textbf{Internal validity} concerns factors, internal to our study, that could have influenced the results. 
A threat is related to the identification of SATD candidates. The keyword-based method employed~\cite{rantala2020prevalence} ensures high accuracy, yet it might bring some selection in the sample, \eg by excluding certain types of SATD because they are not tagged with given keywords. At the same time, given the completely different domain than conventional SATD, an ML-based detection might have introduced more noise in the analysis and still not ensured enough coverage.

\textbf{Conclusion validity} threats mainly concern the representativeness of the sample considered for the analysis. On the one hand, we have sampled a statistically significant number of SATD candidates. On the other hand, such a sample has been extracted on a fairly limited number of projects. 

\textbf{External validity} threats concern the generalizability of our findings. There are three generalizability limitations. First, the study has been limited to Python and the dependents of TensorFlow and PyTorch. While this decision has been justified in \secref{sec:design} given the popularity of Python and both DL frameworks, it is possible that projects developed with other pieces of technology may exhibit different forms of SATD. Second, given the qualitative nature of the study, the analysis has been conducted on a relatively limited number of projects (54), and therefore the conclusions may not generalize beyond such projects. Third, it might be worthwhile to study DL SATD in the closed-source, by employing different empirical methods, \eg surveys or interviews.


\section{Related Work} \label{sec:related}
Considering the scope of our work, we focus on three distinct aspects: (i) TD and SATD in conventional software, (ii) studies on bugs in ML and DL software, and (iii) SATD in ML software.

\subsection{Conventional TD and SATD}
In recent years, the research community has extensively explored Technical Debt as a communication medium among developers and managers to address development issues~\cite{DBLP:conf/icsm/2014mtd, brown2010managing, kruchten2013technical, seaman2011measuring}. Lim \etal~\cite{lim2012balancing} emphasized the intentional introduction of TD, while Ernst \etal~\cite{ernst2015measure} and Zazworka \etal~\cite{zazworka2011investigating} stressed the importance of TD awareness for effective management and mitigation of its negative impact on software quality. 

In their seminal work, Potdar and Shihab~\cite{potdar2014exploratory} found that developers tend to ``self-admit'' TD (SATD) through source code comments. Based on this finding, da Silva Maldonado and Shihab~\cite{maldonado2015detecting} properly categorized different types of SATD, \ie defect, design, documentation, requirement, and test debts, further refined by Bavota and Russo \cite{Bavota2016MSR}. Other work looked at commonalities and differences between SATD within open-source and industry~\cite{zampetti2021self}, \eg industrial developers are less prone to admit TD because of organizational guidelines and career concerns. 


Source code comments are not the only means developers use to trace the presence of TD in the developed code. Xavier \etal~\cite{xavier2020beyond} studied SATD in issue trackers, showing that only a small portion of them (29\%) can be traced to source code. 

Researchers also put effort into properly detecting and categorizing SATD. For instance, Rantala \etal~\cite{rantala2020prevalence} analyzed SATD comments with a detector for Keyword-Labeled SATD, while Ren \etal~\cite{DBLP:journals/tosem/RenXXLWG19} proposed an approach relying on a Convolutional Neural Network to classify source code comments as SATD or non-SATD, as well as to identify key phrases and patterns in code comments that are most relevant to SATD.  


In our work, we focus on SATD categories that are different, and complementary to conventional SATD. As we showed, even SATD related to the need for ``design for change'' pertains to specific changes concerning model deployment on hardware or DL pipeline refactoring.

\subsection{Studies of bugs in ML and DL Software}

Different authors studied bugs in ML and DL frameworks and applications. Zhang \etal~\cite{zhang2018empirical} conducted an empirical study to examine the root causes and symptoms of 175 TensorFlow bugs, as well as how such bugs have been detected and located, highlighting five common strategies for bug detection and localization. Jia \etal~\cite{jia2021symptoms} looked deeper at TensorFlow bugs, analyzing their symptoms and causes, distribution across various components of the library, and repair models. Their findings emphasized the significance of causes over symptoms, as well as commonalities between bugs in traditional software and those in TensorFlow. On the same line, Yang \etal~\cite{yang2022comprehensive} studied bugs in DL frameworks at the source code level. They sampled 1,127 bugs from eight DL frameworks, and manually labeled them according to bug type, root causes, and symptoms. 

Humbatova \etal~\cite{humbatova2020taxonomy} developed a taxonomy of faults in DL systems by manually analyzing 1,059 open-source artifacts and validated it through a survey. The obtained taxonomy featured five primary categories and  375 instances of 92 distinct failure types.

Tambon \etal~\cite{tambon2024silent} conducted an empirical study investigating silent bugs in TensorFlow, categorizing them into 7 scenarios and 4 impact levels based on their effects on DL systems. A survey with 103 TensorFlow users underscored the significant impact of silent bugs in DL frameworks, largely attributed to the stochastic nature of such systems.

Ho \etal~\cite{ho2023empirical} analyzed 194 bugs from the PyTorch framework, investigating their root causes, symptoms, and repair patterns. The findings have been compared with bugs occurring in TensorFlow~\cite{jia2021symptoms}, highlighting both similarities regarding the bug’s root causes, symptoms, and repair patterns, and differences in terms of which components are prone to bugs.

Going deeper into bugs occurring in ML-intensive systems, Morovati \etal~\cite{morovati2024bug} manually examined 386 issues from ML-based systems employing  TensorFlow, PyTorch, and Keras. The findings revealed that ML components exhibit higher error rates compared to non-ML ones, as well as ML-related bugs require more effort to be fixed than conventional ones.



All the aforementioned pieces of research studied bugs in DL frameworks and/or systems. As we have shown in \secref{sec:results}, and discussed more in detail in \secref{sec:dlbugs}, one key difference between DL bugs and DL (self-admitted) TD is that the former may cause crashes or functional errors in common system usages, whereas the latter concern sub-optimal solutions, maintainability problems or, at most, errors occurring in corner-case circumstances.

\subsection{SATD in ML Software}

The most related work to ours is by Obrien \etal~\cite{obrien202223} which analyzes SATD occurring in  ML-based software, proposing a classification scheme accounting for both conventional and ML-specific SATD. While their scope is almost similar to ours, their dataset only accounts for (shallow) ML-based applications while we considered the most forked projects relying on TensorFlow or PyTorch. Furthermore, while our taxonomy has been constructed by considering as dimensions infrastructure and DL lifecycle, including also SATD dealing with inappropriate configuration and setting of the hardware devices to be used for training and inference, their taxonomy considers as dimensions quality aspects such as awareness, modularity, readability, and performance. Last, but not least, there are only 8 overlapping categories across the two taxonomies---the ones related to configurable options and data dependency---highlighting the complementarity between them. 

Bhatia \etal~\cite{bhatia2023empirical} investigated the presence of SATD in 318 open-source ML projects across five domains, and compared them with 318 non-ML projects, in terms of their nature and, also, analyzed their survival over time. Their findings revealed that SATD in ML projects mostly pertains to data pre-processing and model generation. 

Liu \etal~\cite{liu2020using} investigated SATD in the seven most popular open-source DL frameworks. They found a high prevalence of SATD in all studied frameworks, with design debts being the most common. In follow-up work, Liu \etal~\cite{liu2021exploratory} studied the removal of SATD in the same DL frameworks. They found that design debts were most reported and removed quickly, while documentation and defect debts were least self-removed. 
While the studies by Liu \etal~\cite{liu2020using,liu2021exploratory} focused on conventional SATD, we elicited a taxonomy of DL-specific SATD.

Finally, Bogner \etal~\cite{bogner2021characterizing} conducted a systematic mapping study to characterize TD and antipatterns in AI-based systems.
Their study led to the identification of four broad categories of AI TD  (data, model, configuration, and ethical debt). Additionally, they created a catalog of 72 antipatterns, primarily related to data and model, and identified 46 solutions to reduce the accumulation of TD.  Our study has a different setting---analysis of code comments instead of mapping study---relying on actual SATD instances. For this reason, we were able to identify concrete instances of SATD and low-level, code-related problems.


\section{Conclusion and Future Work} \label{sec:conclusion}

The implementation of Deep Learning (DL)-intensive software requires coping with several challenges, including the proper usage of software and hardware infrastructure, as well as the proper creation, tuning, training, and usage of DL models. Sometimes, such challenges may lead to bugs. In other cases, developers decide to leave with sub-optimal solutions, introducing DL-specific Technical Debt (TD).

This paper qualitatively investigates DL-specific Self-Admitted Technical Debt (SATD), \ie SATD occurring in DL-intensive systems, and not related to conventional problems such as maintainability, understandability, or incomplete features. To this aim, we analyzed a statistically significant sample of \sampleSize SATD comments found in \numProjects projects dependent on TensorFlow or PyTorch.

As a result, we elicited a taxonomy of \leaves DL-specific SATD types, organized into \categories categories, in turn further grouped into two high-level categories related to Infrastructure and DL Life-Cycle. We discussed the implications of the found SATDs, as well as the differences with previously-studied DL bugs, and the inadequacy of popular Python static analysis tools to cope with such SATDs.

In future work, we plan to complement this study with further investigations with practitioners, and to develop approaches able to detect and repay DL-specific SATD.

\section*{Acknowledgments}
Massimiliano Di Penta acknowledges the Italian PRIN 2020 Project EMELIOT ``Engineered MachinE Learning-intensive IoT systems'', ID 2020W3A5FY. Federica Pepe is partially funded by the PNRR DM 352/2022 Italian Grant for Ph.D. scholarships. Fiorella Zampetti is partially funded by the PON DM 1062/2021.

\balance
\bibliographystyle{IEEEtranS}
\bibliography{main}

\end{document}